\documentclass[12pt,preprint]{aastex}
\usepackage{natbib}
\usepackage{subfigure}
\usepackage{times}
\begin{document}

\title{On the Nature of the TeV Emission from the Supernova Remnant SN 1006}

\author{Miguel Araya\altaffilmark{1} \& Francisco Frutos}
\affil{Space Research Centre (CINESPA) \\
University of Costa Rica \\
San Jos\'e 2060, Costa Rica}

\altaffiltext{1}{miguel.araya@ucr.ac.cr}

\slugcomment{Submitted to MNRAS on 04/09/2012}

\begin{abstract}
We present a model for the non-thermal emission from the historical supernova remnant SN 1006. We constrain the synchrotron parameters of the model with archival radio and hard X-ray data. Our stationary emission model includes two populations of electrons, which is justified by multi-frequency images of the object. From the set of parameters that predict the correct synchrotron flux we select those which are able to account, either partly or entirely, for the gamma-ray emission of the source as seen by HESS. We use the results from this model as well as the latest constraints imposed by the \emph{Fermi} observatory and conclude that the TeV emission cannot be accounted for by $\pi^0$ decay from high-energy ions with a single power-law distribution, of the form $dN_{\mbox{\tiny proton}}/dE_p \propto E_p^{-s}$, and $s\gtrsim2$.
\end{abstract}


\section{Introduction}
Supernova remnants (SNRs) are thought to be the main source of galactic cosmic rays, high-energy ($E_p\lesssim 10^{15}$ eV) particles (mostly protons) that populate the galaxy, and whose distribution is a power-law. SNRs are energetically favorable and the Fermi mechanism is the natural process by which these objects are able to inject part of their energy into charged particles \citep{bel78,bla87}. Such diffusive shock acceleration (DSA) could account for the observed shape of the cosmic ray distribution.

A direct detection of high-energy protons and nuclei at the sources is hard to obtain, however, and the preferable mechanism to accomplish this task is the detection of MeV--TeV $\gamma$-ray emission from the decay of neutral pions resulting from inelastic collisions between high-energy ions and ambient ions (the so-called hadronic $\gamma$-ray emission). This hadronic signal then has to be separated from (mostly MeV) radiation from the galactic background as well as from (MeV--TeV) radiation from high-energy electrons in SNRs (the so-called leptonic $\gamma$-ray emission).

High-energy electrons were first detected in SNRs through their non-thermal radio spectrum. This radiation is produced when accelerated electrons move in the presence of magnetic fields, which result either from shock compression of the interstellar field or from amplification by cosmic ray instabilities \citep[see][for a recent review]{schure2012}. These electrons typically have energies in the GeV range.

The presence of TeV electrons in young SNRs has been confirmed by X-ray observations. High-resolution observations of young SNRs (e.g., taken by the \emph{Chandra} X-Ray Observatory) show synchrotron emission associated with the forward shock \citep[e.g.,][]{got01,ber02,hwa02,rho02,lon03,vink03,ber04,araya2010}.

The $\gamma$-ray leptonic emission results from inverse-Compton (IC) up-scattering of background photon fields by the synchrotron-emitting electrons and from bremsstrahlung emission from collisions between these electrons and ambient plasma. Leptonic and hadronic high-energy emission could in principle be distinguished based on their relative flux normalization as well as the intrinsic differences in their spectra. However, due to the limited knowledge on the physical parameters of some SNRs, the model degeneracy and limited observations, the identification of hadronic signatures in their non-thermal spectra can be quite challenging.

Spectral energy distributions (SEDs) of SNRs interacting with molecular clouds (which are typically old, with an age $> 10^3$ years), such as W28, W41, W49B, W51C, IC 443, G8.7-0.1 and G359.1-0.5, seem to favor a hadronic origin for the emission \citep{abdo2010b,mehault2011,abdo2010c,abdo2009,abdo2010a,ajello2012,hui2011}, while other recent studies show that the SEDs of some young SNRs with a hard GeV spectrum are probably leptonic-dominated, such as RX J1713.7-3946 \citep{abdo2011a} and RX J0852.0-4622 \citep{tanaka2011}, and have spectral slopes at GeV energies that are consistent with the corresponding IC spectra expected from synchrotron-emitting electrons.

Other young objects, such as the famous historical SNRs Cassiopeia A \citep{arayacui2010,abdo2010d} and Tycho's SNR \citep{mythesis2011} show a soft GeV spectrum which, it might be argued, favors the hadronic scenario \citep[see][however, for an alternative explanation for the emission from Tycho's SNR]{atoyan2011}. Recently, an attempt to unify the properties of $\gamma$-ray emitting SNRs showed that as the density of the environment of the SNR increases, so does the brightness of the hadronic component \citep{yuan2012}.

Regardless of the nature of the high-energy photon emission, it seems clear that the spectral shape below and around hundreds of MeVs is fundamental for its identification. The recently-launched \emph{Fermi} satellite \citep{atw09} has allowed probing this region of the spectrum, although the increasing PSF of its Large Area Telescope (LAT) at these energies and the high galactic background levels may complicate the analysis for some sources.

Here we present a model for the non-thermal emission from SN 1006, the remnant of a type Ia supernova reported by Chinese and Arab astronomers in 1006 \citep{stephenson2002}. The SNR is located far from the galactic center (galactic latitude $\sim 14.5\,^{\circ}$), in a ``clean'' environment with low density \citep[$\sim 0.085$ cm$^{-3}$, see][]{katsuda2009}, at an estimated distance of 2.2 kpc from Earth \citep{winkler2003}. We adopt these values here and develop a two-zone leptonic model to find the sets of reasonable parameters that account for the synchrotron emission observed and are also consistent with the $\gamma$-ray flux as seen by the H.E.S.S. atmospheric Cherenkov telescopes \citep{acero2010}.

As fundamental part of the data used, we included the latest constraints on GeV observations of the source, obtained through analysis of 3.5 years of \emph{Fermi} LAT data. Although the analysis shows that the source is not yet significantly detected at GeV energies, the established upper limit for the emission is used to constrain the SED models.

The results of our simplified model indicate that the TeV flux can only be accounted for by leptonic emission from high-energy electrons in a relatively low ($\sim 30\,\mu$G) magnetic field, hadronic emission from a hard power-law particle distribution,  or by a mixed leptonic-hadronic scenario. In the later cases, hadronic emission from ions with a soft power-law distribution with index higher than 2 can be ruled out as the main cause of TeV photons.

\section{Data}
The SED of SN 1006 contains archival radio observations from different instruments, and were taken from Dyer et al. \citep{dyer2009}. The radio flux density can be described by a power-law with an index of $\sim -0.57$, which can be accounted for by optically-thin synchrotron radiation from a particle distribution with index $-2.14$.

The X-ray fluxes of the source were obtained from observations taken by the Proportional Counter Array (PCA) onboard RXTE on 1996 February. The procedure followed for reducing the data is standard for PCA data, the HEASOFT package version 6.11.1 was used. Only events from the top layer of PCU2 were used for analysis in the interval from 3.5 keV to 25 keV, and good time intervals were generated by filtering out events during South Atlantic Anomaly passages, times when SN 1006 was less than $10^{\circ}$ above the limb of the Earth and the pointing direction of the detectors is more than $0.^{\circ}02$ from the nominal pointing direction. The PCA background spectrum for SN 1006 was calculated using version 3.8 of the FTOOL \texttt{pcabackest}\footnote{For more information about FTOOLS, see http://heasarc.gsfc.nasa.gov/docs/software/ftools .} and the background model \emph{pca\_bkgd\_cmfaintl7\_eMv20051128.mdl}. The background-subtracted spectra from the source for each independent observation were combined with the FTOOL \texttt{addspec} and appropriate responses were generated with \texttt{addrmf}.

The resulting X-ray spectrum was binned to contain 1000 counts per bin and fitted in XSPEC version 12.7.0\footnote{For more information about XSPEC, see http://heasarc.gsfc.nasa.gov/docs/xanadu/xspec/index.html .} with an absorbed power-law model after adding $1\%$ systematic error. The column density cannot be constrained and is fixed at $N_H =7\times10^{20}$ cm$^{-2}$, according to the literature \citep[][and also see the FTOOL \texttt{nh} which gives archival galactic column density values]{kalberla2005}.

The fit obtained has a reduced $\chi^2$ of 1.08, and the resulting values for the photon index and unabsorbed flux (with $90\%$-confidence errors) are $2.83^{+0.13}_{-0.12}$ and $(2.08\pm{0.11})\times10^{-11}$ ergs/cm$^2$/s (in the energy band $3.5-25$ keV), respectively. The resulting unfolded spectrum, corrected for absorption using standard photoelectric cross section values below 10 keV \citep{morrison1983}\footnote{The spectral absorption factor is $e^{-N_H\sigma(E)}$, where $\sigma(E)$ is the cross section for a photon energy $E$.}, is shown in Fig. \ref{fig1} with the rest of the SED.

TeV $\gamma$-ray observations of SN 1006 published by the H.E.S.S. Collaboration \citep{acero2010} and the latest constraints imposed by the \emph{Fermi} LAT obtained here complete the high-energy SED of the source.

\subsection{Analysis of \emph{Fermi} LAT data}
We analyzed the \emph{Fermi} LAT data gathered between 04 August 2008 and 16 January 2012. The analysis was performed with the LAT Science Tools package \footnote{See http://fermi.gsfc.nasa.gov/ssc} version v9r23p1. The cuts and selection criteria applied are those recommended by the Fermi Science Support Center for binned likelihood analysis\footnote{See http://fermi.gsfc.nasa.gov/ssc/data/analysis/scitools/binned\_likelihood\_tutorial.html .}, which is generally used for large data sets such as this one. We selected diffuse events \citep[see][]{atw09} between 500 MeV and 100 GeV within a region of interest (RoI) of $30^{\circ}$ around the position of the source, RA (J2000)= 15$^h$02$^m$22.1$^s$, Dec (J2000)= -42$^\circ$5$'$49$''$, and a maximum zenith angle cut of $100^\circ$, in order to avoid the background from the limb of the Earth, with the tool \texttt{gtselect}. Appropriate livetime calculation with the tool \texttt{gtmktime} is performed, excluding time intervals where the zenith cut intersects the RoI.

In order to account for all the photons from background sources, an `xml' file is created which contains information about source positions an spectra as published in the LAT 2-year Source Catalog \citep{abdo2011b}. The resulting file contains 260 point sources and 2 extended sources (known as Cen A and MSH 15-52), which are modeled according to the emission templates provided by the Fermi Collaboration\footnote{Included within the software distribution, or downloaded at \\ http://fermi.gsfc.nasa.gov/ssc/data/access/lat/2yr\_catalog/gll\_psc\_v07\_templates.tgz .}. Modeling such a large region of the sky is necessary due to the large PSF of the telescope. The size of the region and the time span of the observation make the analysis computationally extensive. However, the source is located relatively far from the galactic center which reduces the uncertainties in the galactic background estimation.

The current instrument response functions ($P7SOURCE\_V6$) are used throughout the analysis, as well as the latest galactic and extragalactic diffuse background components (as specified in the files \emph{gal\_2yearp7v6\_v0.fits} and \emph{iso\_p7v6source.txt}, respectively). The spectral parameters of the catalogued sources beyond 30 degrees from the position of SN 1006 are kept frozen during the fit, which is performed several times with the use of the optimizer NEWMINUIT until convergence is achieved.

The diameter of the X-ray shell of the remnant, as seen for instance by \emph{Chandra} \citep{lon03}, is about 30 arcmin. The averaged 95\%-containment angle for (front+back) events for the LAT is about $0.8\,^{\circ}$ at the highest energies \citep{atw09} where the resolution is also the highest. This would make SN 1006 hard to resolve by the \emph{Fermi} LAT\footnote{An extended source hypothesis was tested in the likelihood analysis as well, following the emission profile for the source reported from the Sydney University Molonglo Sky Survey \citep{bock1999} at 843 MHz, and the results do not change significantly with respect to the point-source model.}. A point source hypothesis for SN 1006 was then used to derive the main flux constraints presented in this paper.

\subsubsection{Results of the LAT data analysis}
The emission from SN 1006 at GeV energies is not significant. The resulting value of the test-statistics\footnote{The modeling of \emph{Fermi} LAT data involves a likelihood fit where the significance of a source is quantified with the value of its TS \citep[see][for details]{mat96}.} (TS) for a point-source at the position of SN 1006 with free spectral parameters is only 7.5, which corresponds to a detection significance of only about $\sqrt{\mbox{TS}} \sim 2.7\sigma$ above the background.

If the $\gamma$-ray emission from the source is assumed to follow a power-law distribution, $dN_\gamma/dE_\gamma \, \propto \, E_\gamma^{-\Gamma}$ (in MeV$^{-1}$ cm$^{-2}$ s$^{-1}$, with $\Gamma$ the \emph{photon} index), with $\Gamma=2.14$, as would be expected from emission of hadronic origin from a high-energy proton population with a distribution similar to the one observed for the high-energy electrons, $dN_{\mbox{\tiny proton}}/dE_p\, \propto \, E_{p}^{-2.14}$ (with $E_p$ the proton energy), the TS value decreases to about 4.9, and the estimated $99\%$-confidence level upper limit on the integrated source flux in the energy interval of the analysis ($0.5-100$ GeV) under this assumption is $1.32\times10^{-9}$ photons/cm$^2$/s.

The TS values obtained for different assumptions on the index of the power-law photon distribution by a point source at the position of SN 1006 are found to decrease with higher photon index (i.e., for softer $\gamma$-ray spectra). For example, the TS value obtained for a photon index $\Gamma=2.0$ is 5.7, and for $\Gamma=1.6$, 7.2. This latter value corresponds to the slope of the photon distribution (i.e., in photons MeV$^{-1}$ cm$^{-2}$ s$^{-1}$) expected for GeV emission resulting from IC scattering by an electron population with index 2.14, as the one observed in SN 1006. Table \ref{table1} shows a summary of these results. The upper limit obtained for $\Gamma=1.6$ is shown in the SED (Fig. \ref{fig1}) and is used to constrain our leptonic model, described in the next Section.

\section{Model}
\subsection{Scales of emission regions and particle distributions}\label{imagesSection}
The observed radio and X-ray synchrotron emission is accounted for with a two-population stationary model. The diffuse radio electrons, which we denote as Zone 1, are uniformly distributed over a volume $V_1$ and higher energy electrons in Zone 2 associated with the X-ray rim near the forward shock region, with a volume $V_2$.

From the radio and X-ray images of the source shown by Cassam-Chena\"i et al. \citep{cassam2008}, we estimated that $V_2$ is about $11\%$ of the volume of the SNR (or $V_2 \sim 1.4\times10^{55}$ cm$^3$, for an spherical SNR of radius $9.5$ pc), and the volume fraction of Zone 1 is about $62\%$ of the volume of SN 1006 ($V_1 \sim 7.6\times10^{55}$ cm$^3$).

Next, the non-thermal emission from these two populations is calculated, using power-law particle distributions in the two Zones, as usually done to account for optically thin synchrotron radiation. In fact, in order to account for the X-ray data, a broken power-law particle distribution is used in Zone 2, and similarly in Zone 1 as required for the high-energy $\gamma$-ray data (see below).

The synchrotron emission from Zone 1 dominates at radio frequencies, while Zone 2 is brightest at X-ray energies. Below the break energy, the power-law index for both particle distributions is fixed at 2.14, which is required by the radio spectrum.

We calculated the resulting $\gamma$-ray emission from these two populations. It included inverse Compton scattering of cosmic microwave background photons (IC/CMB) and non-thermal bremsstrahlung (NB). For the latter we considered interactions of electrons with other electrons, protons, and fully ionized helium. We used an abundance for helium of 0.1 that of hydrogen. The contributions from synchrotron self-Compton scattering are entirely negligible.

The NB flux is proportional to the density of the target plasma, which for Zone 2 can be taken as four times the local ISM density, $n^{\mbox{\tiny Zone 2}}_H = 0.3$ cm$^{-3}$, as is standard for a strong shock \citep[e.g.][]{drury1983}, and for Zone 1 it can be estimated as the mass of the ejecta (1.4 M$_{\odot}$) plus a comparable mass of shocked ISM material distributed in the volume of the SNR, $n^{\mbox{\tiny Zone 1}}_H = 0.04$ cm$^{-3}$. These values result in negligible NB emission for all reasonable values of the model parameters.

The contribution to the $\gamma$-ray emission from Zone 2 is negligible for all the reasonable values of the magnetic field in this Zone (see Section \ref{magneticfield}). The particle indices beyond the break are $2.8$ for Zone 1 and $2.9$ for Zone 2, necessary to reproduce the $\gamma$-ray and X-ray observations, respectively. It is beyond the scope of this paper to speculate on the nature or cause of the particle breaks, and we limit ourselves to point out the need of such spectral distributions if the emission is to be accounted for by leptonic processes alone.

The two particle distributions in our main leptonic emission model (Fig. \ref{fig1}) can be reasonably well described by the following functions $$\frac{dN_e}{dE_e dV}^{\mbox{\tiny Zone 1}} = (2.06\times 10^{-9}\,\mbox{GeV$^{-1}$ cm$^{-3}$}) \left(\frac{E_e}{\mbox{1 GeV}}\right)^{-2.14} \left(1+\left(\frac{E_e}{\gamma^{\mbox{\tiny Zone 1}}_{br}m_ec^2}\right)^2\right)^{-0.33}\,\, \mbox{and}$$

$$\frac{dN_e}{dE_e dV}^{\mbox{\tiny Zone 2}} = (1.84\times 10^{-10}\,\mbox{GeV$^{-1}$ cm$^{-3}$}) \left(\frac{E_e}{\mbox{1 GeV}}\right)^{-2.14} \left(1+\left(\frac{E_e}{\gamma^{\mbox{\tiny Zone 2}}_{br}m_ec^2}\right)^2\right)^{-0.38},$$ where $m_e$ is the mass of the electron, and $\gamma^{\mbox{\tiny Zone 1}}_{br}=7\times 10^{5}$ and $\gamma^{\mbox{\tiny Zone 2}}_{br}=3\times 10^{6}$ are the break Lorentz factors for Zone 1 and Zone 2 respectively. The corresponding maximum Lorentz factors are $5\times10^{7}$ and $6\times10^{7}$.

\subsection{Magnetic field and energy in electrons}\label{magneticfield}
We adopt a value for the magnetic field in the X-ray rim of SN 1006 (Zone 2) consistent with constraints on the diffusion and acceleration parameters made by Parizot et al. \citep{parizot2006}. Their calculations, obtained assuming uniform, isotropic turbulence, are based on observed SNR X-ray properties such as the synchrotron cutoff energy, the velocity of the expanding shock and the rim thickness.

The values obtained by Parizot et al. \citep{parizot2006} for the field in the downstream region of SN 1006 are $\sim 60 - 100 \,\mu$G, depending on the value of the compression ratio and the energy dependence of the diffusion coefficient used. They used a value for the projected width of the rim that is consistent with other estimates \cite[$l=0.2$ pc, see][]{bamba2003} and a distance to the remnant of 2.2 kpc. Although they used a value for the shock speed of 2900 km/s that is a factor of 1.7 lower than more recent estimates \citep{katsuda2009}, which according to their formula for the downstream magnetic field translates into a field value around 1.2 times larger.

The value adopted here is $80\,\mu$G, which results in an energy content of $W_e^{\mbox{\tiny Zone 2}}= 7.7\times10^{43}$ erg for the corresponding electrons. This value for the magnetic field is also consistent with other values obtained from models applied to high resolution X-ray data \citep{morlino2010}, although considerably lower compared to some nonlinear kinetic models \cite[$\sim 150\,\mu$G, see][]{volk2005}. We explored the effects of using such a range of values for the field in Zone 2 in our model (see below).

With the parameter values described so far, the resulting IC emission from Zone 2 is small, and considerably below the observed H.E.S.S. flux at TeV energies. At this point we consider a leptonic scenario for the very high energy emission from SN 1006. If the $\gamma$-ray emission is attributed to Zone 1, the magnetic field required there should be relatively low, around $10\,\mu$G, resulting in an energy content of $W_e^{\mbox{\tiny Zone 1}}= 4.2\times10^{45}$ erg.

In such a leptonic scenario, the break energy in the particle distribution in Zone 1 should be lower than the corresponding value for Zone 2, in order to account for the observed H.E.S.S. emission at the highest energies without increasing the GeV fluxes, which are constraint by the \emph{Fermi} LAT observations.

The resulting modeled emission as well as the observations are shown in Fig. \ref{fig1}. As can be seen, the exact slope of the H.E.S.S. data is hard to reproduce with the model. This is the case for a wide range of model parameters.

\subsection{On the parameter degeneracy}
If the model parameters for Zone 1 are now fixed to the values described above, the required magnetic field in Zone 2 cannot be lower than about $30\,\mu$G, for a corresponding energy in electrons in this Zone of $5.6\times10^{44}$ erg and a higher maximum Lorentz factor of $1.2\times10^{8}$. The predicted $\gamma$-ray SED flux from Zone 2 for this field would be about 5 times lower than the emission from Zone 1 at most frequencies, but the lower limit on the magnetic field results from fitting the $\gamma$-ray data with the contributions from both Zones. In other words, this range of values is allowed by the uncertainty in the data.

Conversely, if a higher value of the magnetic field in Zone 1 (the radio emitting electrons) is used, the TeV fluxes could instead be attributed to Zone 2, where the field should be lower. It was found that for a magnetic field in Zone 1 of $40\,\mu$G, the required field in Zone 2 should be about $17\,\mu$G. Also, the break and maximum Lorentz factors would be higher than the values used above, and both electron populations would have comparable total energies. However, this field in Zone 2 is not realistic for the observed widths of the filaments. For this reason, we adopted the parameters described in Sections \ref{imagesSection} and \ref{magneticfield} for the rest of the analysis. We refer to this particular parameter selection as our main leptonic model.

\subsection{Hadronic emission}
We studied the effect of adding a component of $\gamma$-ray emission from hadronic interactions to the modeled non-thermal SED of SN 1006. The hadronic emission is calculated as in Kamae et al. \cite{kamae2006} for power-law particle distributions with various slopes. Figures \ref{fig2} and \ref{fig3} show the corresponding models obtained for different hadronic emission levels and particle indices.

When assessing the possible amount of hadronic flux, the parameters for Zone 2 were fixed for all the models to the values shown in Sections \ref{imagesSection} and \ref{magneticfield}, but the amount of IC emission from Zone 1 is varied to maintain consistency with the data as required.

Figure \ref{fig2} shows the result of emission models with a steep hadronic population described by a single power-law with index 2.14. This value also corresponds to the index of the distributions of the leptonic components. Such a value for the index of the proton distribution is typical of hadronic $\gamma$-ray emission used in current SED models, both of SNRs and SNRs interacting with molecular clouds \citep{arayacui2010,abdo2010a,abdo2010c,abdo2010e,mythesis2011,giordano2012}.

For the case of the hadronic-dominated scenario, the value of the magnetic field in Zone 1 is $40\,\mu$G, resulting in an energy content in electrons of $4.6\times10^{44}$ erg. The shape of the spectrum of the radio-emitting electrons is the same as before, but the required maximum Lorentz factor in Zone 1 is lower than in the leptonic-dominated scenario. The maximum particle energy reached by protons, as required by the H.E.S.S. data, is 90 TeV.

As can be seen in Fig. \ref{fig2}, the hadronic model over-predicts the flux at GeV energies and therefore is not consistent with the limit imposed by observations. In fact, the same conclusion can be obtained for any power-law proton distributions with an index higher than 2.0, due to the shape of the resulting spectra \citep[the $\gamma$-ray flux from hadronic interactions maps the particle spectrum directly, see, e.g., Kamae et al.][]{kamae2006}. Therefore, hadronic TeV emission from a cosmic ray population with index greater than 2.0 at all energies at the level observed by H.E.S.S. is discarded for SN 1006.

For a proton distribution with index 2.14, Fig. \ref{fig2} also shows a scenario where the hadronic emission is the maximum possible that is also compatible with observations. This is achieved with a magnetic field and energy in Zone 1 of $12\,\mu$G and $3.0\times10^{45}$ erg, respectively.

Similar models can be obtained for a different index for the proton distribution. For a proton distribution with index 2.0 (the value obtained from DSA calculations for strong shocks in the test-particle limit), models are shown in Fig. \ref{fig3} for the case where the emission is dominated by hadronic processes (and which is obtained by suppressing the IC emission with a magnetic field in Zone 1 of $50\,\mu$G and an energy in electrons of $3.0\times10^{44}$ erg) and for a case where the leptonic and hadronic fluxes are comparable at TeV energies (with $B^{\mbox{\tiny Zone 1}}=16\,\mu$G and an energy in Zone 1 of $2.1\times10^{45}$ erg), which we argue is more consistent with the data (particularly with the LAT upper limit). In these two cases, the proton cosmic ray spectrum extends up to 45 TeV.

\section{Discussion and Conclusions}
From the simple model presented here, we can conclude that if a value for the magnetic field in Zone 2 is used which would reproduces the widths of X-ray filaments near the forward shock of SN 1006 ($\sim 80\,\mu$G), then a leptonic scenario for the $\gamma$-ray emission implies that high-energy photons should originate mainly from radio-emitting electrons in Zone 1. In this scenario, the required electron spectrum is described by a broken power-law with indices 2.14 (as required by the radio data) and 2.8, and a break around a particle energy of 360 GeV, as required by the H.E.S.S. observations.

For a lower magnetic field in Zone 2 ($20 - 30\,\mu$G), the level of the observed $\gamma$-ray flux can be attributed mainly to this Zone. We cannot rule out either scenario with our current knowledge of the magnetic field. The shape of the TeV spectrum in our model could be improved with a combination of such relatively low magnetic field values in both Zones.

The more interesting result obtained from our analysis of the observations is related to the constraints imposed by cumulative observations from the \emph{Fermi} LAT, which allow us to conclude that the TeV emission from SN 1006 cannot be produced by $\pi^0$ decay from a hadronic population with a ``soft'' power-law distribution (i.e., with an index $\sim 2.0$ or higher), such as the ones typically used in similar phenomenological models of broad-band SEDs of SNRs \citep[e.g.,][]{arayacui2010,abdo2010a,abdo2010c,abdo2010e,mythesis2011,giordano2012}, but they are consistent with a mixed leptonic-hadronic (or leptonic-dominated) model for the case where the index of the high-energy ion component is around 2, or with a pure leptonic model for softer cosmic ray distributions, as is seen in Figs. \ref{fig2} and \ref{fig3}. Of course, the steeper the hadronic distribution, the lower its contribution to the TeV $\gamma$-ray flux should be. We cannot rule out a hard hadronic radiation spectrum (i.e., with a particle index below $2$).

It is important to stress the fact that our model only rules out hadronic distributions that are single steep power-laws at all energies as the main cause of the TeV emission from the source, and which differ from the ones predicted by some nonlinear models with efficient acceleration \citep[e.g.,][]{chevalier1983,malkov1999,blondin2001} which show spectral hardening at high energies. However, such proton distributions are not seen in other high-efficiency acceleration models \citep[e.g.,][]{ellison2010}.

\begin{figure}[htp]
\begin{center}
\includegraphics[width=8cm,height=5cm]{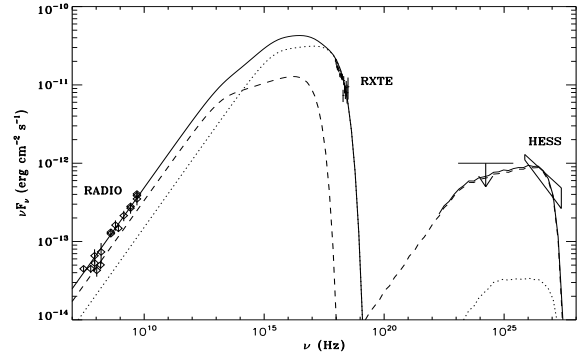}
\caption{SED of SN 1006 and leptonic emission model. The model includes synchrotron and IC/CMB emission from Zone 1 (dashed lines) and from Zone 2 (dotted lines), and total emission (solid line). The parameters of the fit are reported in Sections \ref{imagesSection} and \ref{magneticfield}.}
\label{fig1}
\end{center}
\end{figure}

\begin{figure}[h]
\begin{center}
\subfigure[ ]{\includegraphics[width=8cm,height=5cm]{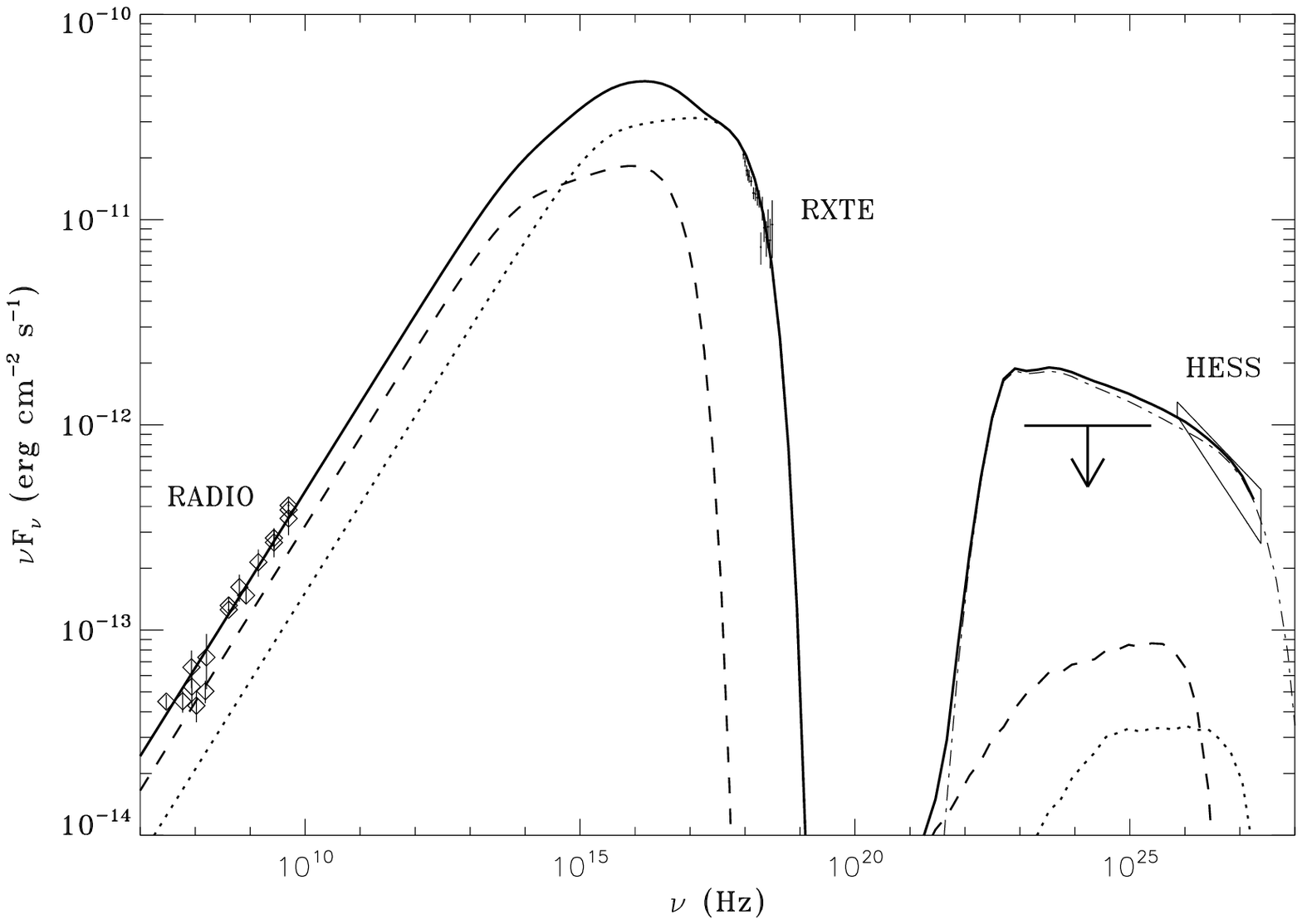}}
\subfigure[ ]{\includegraphics[width=8cm,height=5cm]{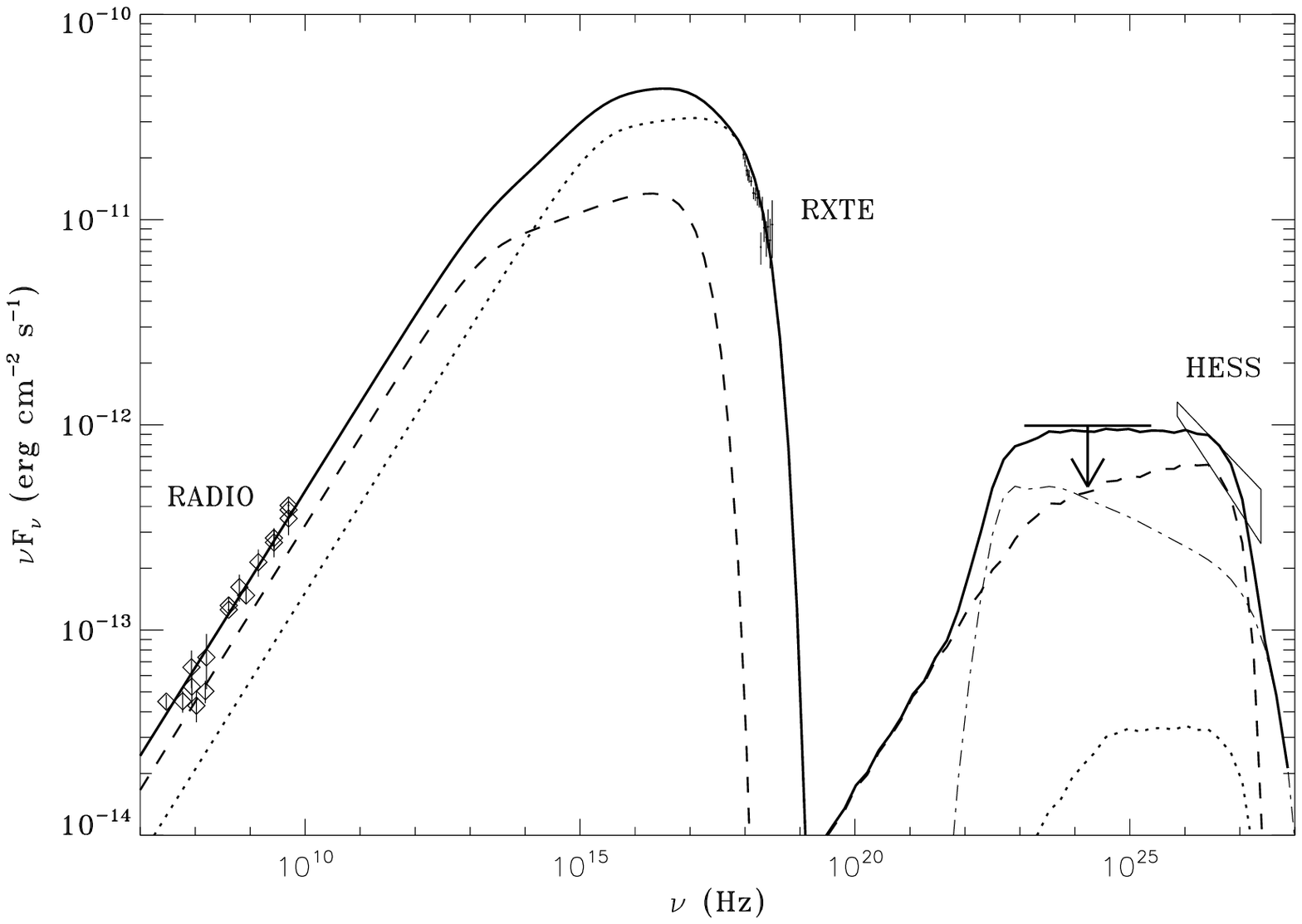}}
\caption{SED of SN 1006 and models with a contribution from a hadronic component produced by a power-law distribution of particles with index 2.14 (dash-dotted line), (a) assuming a hadronic origin for the TeV emission; and (b) showing the maximum hadronic $\gamma$-ray flux allowed by the data.}
\label{fig2}
\end{center}
\end{figure}

\begin{figure}[h]
\begin{center}
\subfigure[ ]{\includegraphics[width=8cm,height=5cm]{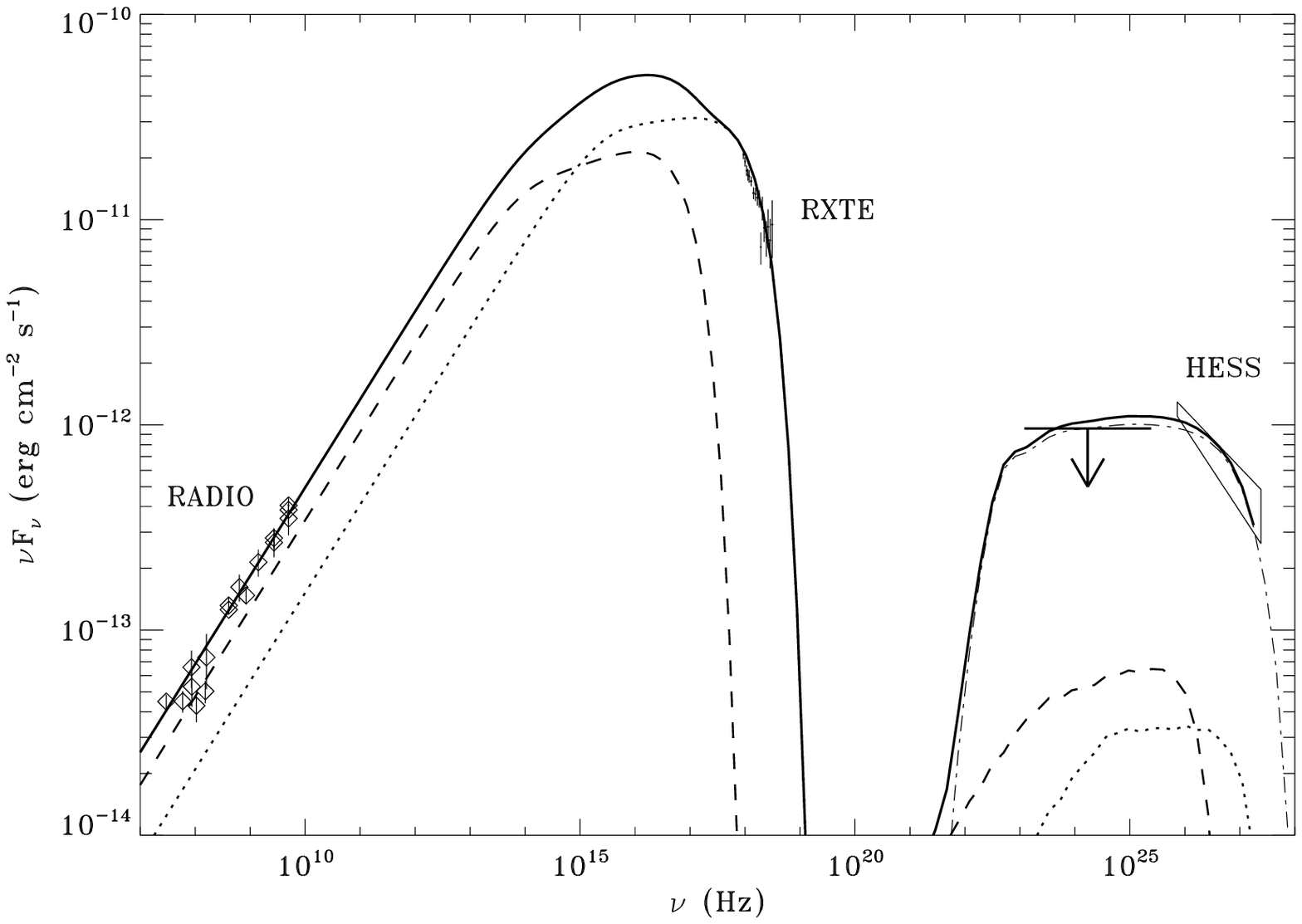}}
\subfigure[ ]{\includegraphics[width=8cm,height=5cm]{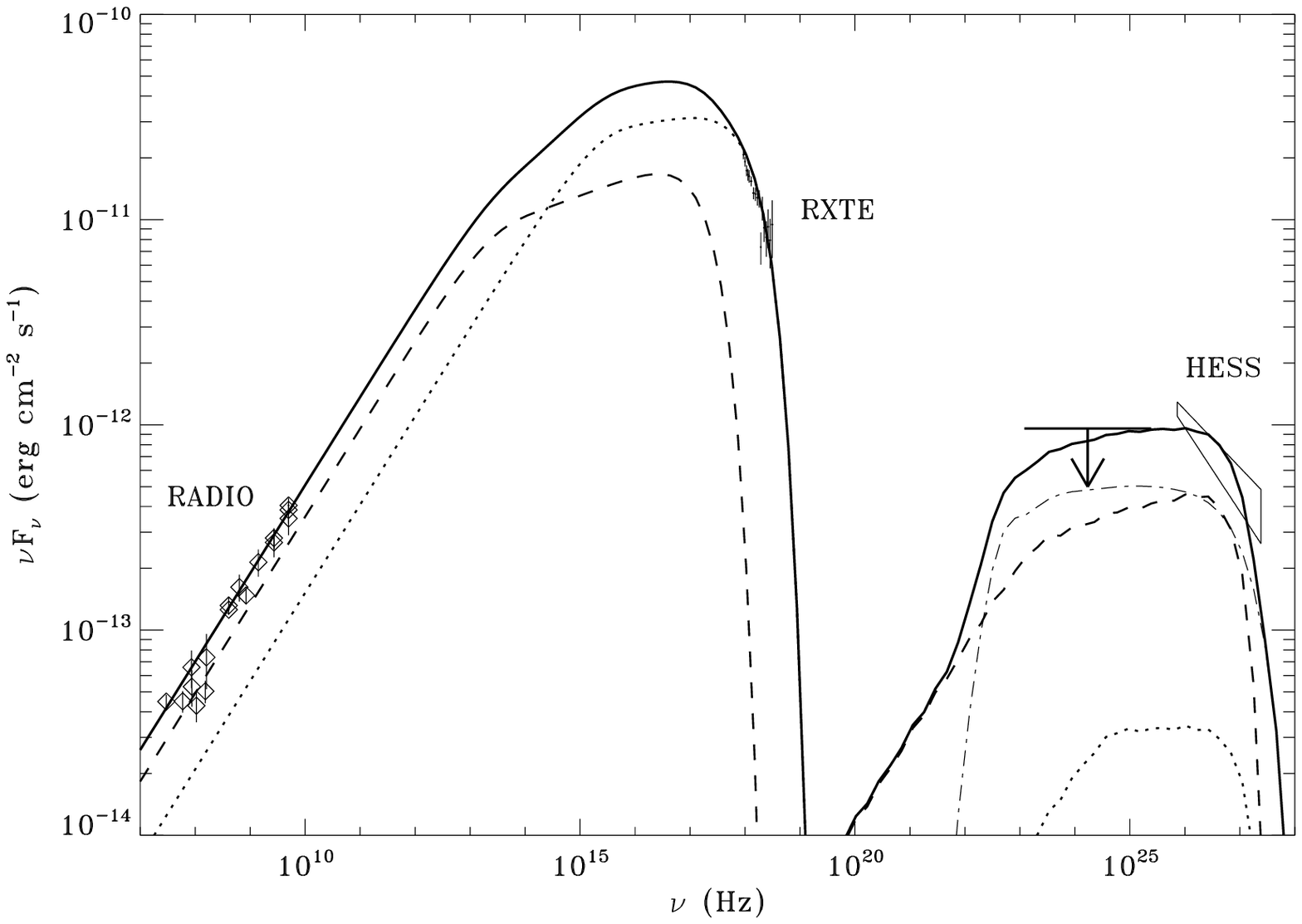}}
\caption{Same as Fig. \ref{fig2} for models with a hadronic component produced by an ion distribution with index 2.0 (dash-dotted line), corresponding to (a) the hadronic-dominated scenario; and (b) the mixed leptonic-hadronic case.}
\label{fig3}
\end{center}
\end{figure}

We also point out that our model is an extremely simple approximation which neglects all aspects of the temporal dynamics as well as nonlinear effects from the backreaction of accelerated particles in the fluid dynamics \citep[see, e.g.,][for a description of spectral curvature seen in the radio spectrum of SN 1006 and possibly resulting from nonlinear dynamics]{allen2008}, MHD turbulence and source photons other than the CMB that could be up-scattered and contribute to the IC flux. In fact, as can be seen in Fig. \ref{fig1}, the shape of the H.E.S.S. spectrum of SN 1006 is hard to reproduce with our model, which is true for any other phenomenological IC--CMB emission model unless a rather odd spectral shape is used for electrons in Zone 1 (e.g., a distribution with a spectral break at low energies and a very steep spectrum, with a power-law index of $\sim 3.7$, above the break energy). A full broadband nonlinear model of the non-thermal emission could perhaps account for the spectral shape at TeV energies, but our main conclusion would still be valid.

Based on our model it can be pointed out, as explained before, that a lower magnetic field in the X-ray emitting region is supported by the H.E.S.S. observations if the nature of the TeV emission is leptonic. The bipolar morphology of the $\gamma$-ray emission seen by H.E.S.S. indicates that it originates in the polar caps of the remnant, and in their proposed leptonic scenario, the flux level is consistent with a field value of $30\, \mu$G as pointed out by the authors \citep{acero2010}. For their hadronic scenario, they find that a field of $120\, \mu$G is consistent with the data, and necessary to ``suppress'' the leptonic emission. Our results from Section \ref{magneticfield} are consistent with their leptonic emission scenario. We have already pointed out that a field value of $30\, \mu$G in Zone 2 of our model can also account for the H.E.S.S. observations.

On the other hand, other authors have derived a higher value for the magnetic field. Using a detailed nonlinear kinetic model for its non-thermal X-ray emission, some estimates of the magnetic field in SN 1006 favor a value of $150\, \mu$G \citep{volk2005}, which then translates into a predicted IC flux that is lower than the $\gamma-$ray flux observed by H.E.S.S., as can also be concluded with our model. This has lead other authors to argue that the TeV emission might be predominantly hadronic \citep{berezhko2009}. We can only point out that if such were the case, then our GeV limit on the emission implies that the hadronic $\gamma$-ray spectrum should be hard at GeV energies, which is our main conclusion, and this is in accordance with the model shown by these authors \citep{berezhko2009}.

The main leptonic scenario for the TeV emission in our two-Zone model presents other difficulties. The break in the particle spectra in Zone 1 which is necessary to reproduce the observations cannot be accounted for by synchrotron cooling alone. The particle break was introduced in the basis of observations. A break in the predicted IC emission would be in general required by the LAT constraints and the break energy itself would be needed to reproduce the observed TeV spectral shape.

With respect to the predicted bremsstrahlung flux, the fact that this emission (which reproduces the particle spectrum directly) from SN 1006 is found to be negligible for all the leptonic parameters shown here means that the spectrum of the source at GeV energies would either be hard and leptonic in nature, hard and hadronic in nature or a relatively flat spectrum of mixed origin. The results of the \emph{Fermi} LAT data analysis shown in Table \ref{table1} seem to indicate that the data favor a hard spectrum for the emission (as shown by the increasing values of the TS for lower photon indices), which would also be consistent with the hadronic model mentioned above \citep{berezhko2009}. This will only be confirmed with a significant GeV detection of the source in the future.

With respect to hadronic emission from a particle population with an index of 2 (Fig. \ref{fig3}), it is seen that the data are more consistent with mixed leptonic and hadronic contributions (e.g., of comparable fluxes at TeV energies) rather than a predominantly hadronic scenario, for which the resulting fluxes are slightly above the GeV upper limit.

Finally, we point out that despite its problems, a leptonic origin of the non-thermal emission from SN 1006 at all energies would be consistent with the fact that other ``SN 1006-like'' SNRs such as RX J1713.7-3946 \citep{koyama1997}, show $\gamma$-ray spectra and shock structure consistent with leptonic emission \citep{ellison2010,abdo2011a}. Also, it has been proposed recently that non-thermal emission from SNRs in low density environments (such as SN 1006) is caused primarily by high-energy leptons \citep{yuan2012}.

We cannot of course conclude that the TeV emission from the source is mainly hadronic or leptonic, but we stress the fact that the data rule out certain hadronic emission models such as the ones that have been used to account for the observations of other young SNRs, such as Cas A \citep{abdo2010b,arayacui2010} and Tycho's SNR \citep{giordano2012}.

\acknowledgments
This research has made use of NASA's Astrophysical Data System. We thank the anonymous referee for useful comments on the manuscript. We also gratefully acknowledge financial support from Universidad de Costa Rica.

\begin{table}
\begin{center}
\caption{\emph{Fermi} LAT Upper Limits on the GeV SED of SN 1006 for Different Assumed Photon Spectral Indices\label{table1}}
\begin{tabular}{|c|c|c|}
\tableline\tableline
Index & $E_\gamma^2\frac{dN_\gamma}{dE\gamma}^{\mbox{\tiny UL}}$ (ergs cm$^{-2}$ s$^{-1}$) & TS\\
\tableline
1.6 & $1.0\times10^{-12}$ $^{a}$ & 7.2\\
2.0 &  $9.6\times10^{-13}$ & 5.7\\
2.14 &  $9.9\times10^{-13}$ $^{b}$ & 4.9\\
\tableline
\end{tabular}
\tablenotetext{a}{Evaluated at an energy of 5.7 GeV.}
\tablenotetext{b}{Evaluated at an energy of 2.1 GeV.}
\end{center}
\end{table}

\clearpage

\end{document}